\documentclass[12pt]{article}
\usepackage{amssymb}
\usepackage{amsmath}
\usepackage{thmsa}

\begin{document}

\author{Emilio Santos \\
Departamento de F\'{i}sica.\\
Universidad de Cantabria. Santander. Spain \\
email: santose@unican.es}
\title{Are Bell's conditions for local realism general enough ?}
\date{March,15,2025}
\maketitle

\begin{abstract}
Bell conditions for local realism are critically revisited. In particular
for optical experiments I criticize Bell's proposed response of detectors
to signals as extremely idealized. More physical conditions are proposed,
whence a realistic local model of an optical experiment is possible which
violates the Clauser-Horne (Bell) inequality. The possibility rests on the
existence of a coincidence-time loophole in the experiments.
\end{abstract}

\tableofcontents

\section{Introduction}

The loophole-free violation of Bell inequalities \cite{Bell64}, \cite{Bell}
in experiments with entangled photon pairs \cite{Shalm}, \cite{Giustina}, 
\cite{BIG}, has been claimed the death by experiments for local realism \cite
{Wiseman}, \cite{Aspect}, \cite{Zeilinger}. Local realism has been defined
by H. Wiseman as the assumption that: ``The world is made up of real stuff,
existing in space and changing only through local interactions. This
local-realism hypothesis is about the most intuitive scientific postulate
imaginable'' \cite{Wiseman}.

The origin of the Bell inequalities may be traced back to the celebrated
article by Einstein, Podolski and Rosen (EPR) \cite{EPR} published 90 years
ago. Analyzing an imaginary but feasible experiment the authors proved that
either \textit{relativistic causality} (locality) is violated or the quantum 
\textit{decription of physical reality} (realism) is incomplete . As is
known Einstein always supported that locality should hold true \cite
{biography}. Thirty years after EPR, Bell pointed out that the conflict is
more dramatic, local realism is not only incompatible with \textit{the
completeness} of quantum mechanics, but with its\textit{\ testable
predictions} too\cite{Bell64}. Indeed Bell derived correlation inequalities,
claimed to be straightforward consequences of local hidden variables
theories (HVT), which are violated by quantum predictions in some instances.
In the years elapsed from Bell's work until today ``local realism'' has
been substituted for ``local HVT'' as a more general and appropriate topic
(but see section 2 below). Hence the dilemma `quantum mechanics vs. local
realism' that has been put to empirical scrutiny via the experimental tests
of Bell inequalities.

Most the academic community believes that the dilemma is closed as a
consequence of the ``loophole-free'' experiments \cite{Aspect}. However both
terms of the alternative are awful, because each term implies the conflict
with one of the two fundamental theories of physics, that is quantum
mechanics and relativity. Therefore searching for vulnerable points is
pertinent, and this is the motivation of the present article.

The experimental tests of Bell inequalities have proved to be very difficult
to perform, as shown by the 50 years delay from Bell's work to the first
``loophole-free'' experiments \cite{Shalm}, \cite{Giustina}. In particular
the constraint that the measurements made by two parties (say Alice and Bob)
should be space-like separated is difficult to achieve with massive
particles, so that it seems almost unavoidable to use entangled photon
pairs. As a consequence most tests of Bell inequalities performed till now
have been optical experiments. However efficient detection of individual
photons is not easy, whence a persistent difficulty for the optical tests
has been the low efficiency of photon counters. The problem was not solved
until recently, after efficiencies greater than 70\% became available.
Closing all loopholes has been therefore a formidable achievement on the
experimental side.

In this paper I do not question that the Bell inequalities have been
violated in experiments. Indeed from the time of Bell's paper \cite
{Bell64} till present day the possible loopholes in the performed
experiments have been carefully scrutinized and the community has arrived at
an agreement about the reliability of the latest experiments. In contrast
the derivation of the Bell inequalities has been scarcely questioned after
the early work by Clauser et al. \cite{CHSH}, \cite{CH}, but I dispute that
all local realistic models should fulfil the inequalities. Analyzing
critically the derivations of Bell inequalities from local realism is the
main purpose of the present paper. In particular I will question in section
2 the generality of Bell's conditions for local realism. The point is
that using a more general formulation of local realism there may be
classical models able to violate the inequalities. The proof is made via
proposing a model that does that, in section 3.1. In section 3.2 I shall
show that, if we accepted Bell formalization (eqs.$\left( \ref{bell}\right) $
and $\left( \ref{bell1}\right) $ below), the predictions of the model fulfil
the Clauser-Horne (Bell-type) inequality. In section 3.3 I will prove that
the model may violate a Bell inequality if the more appropriate
formalization of sections 2.2 and 2.3 is used. Our model is related to the
so-called coincidence-time loophole, see section 2.4 below.

\section{Local realism and Bell inequalities}

\subsection{Bell's conditions for local hidden variables}

In order to critically analyze Bell's work it is convenient to start
mentioning that Bell wrote about local hidden variables (LHV), rather than
about local realism (LR) as Einstein et al made \cite{EPR}. This is relevant
because LHV and LR are quite different concepts in my opinion. I believe
that the Bell inequalities are necessary conditions for LHV, but not for LR.
The point is clarified in the following.

The origin of hidden variables goes back to the early period of quantum
mechanics. Indeed a characteristic trait of quantum physics is that
measurements made on identical systems prepared in the same state may give
different results, at a difference with what happens in the classical
domain. This is currently expressed as a lack of strict causality in the
quantum domain. However there is an alternative that may preserve causality,
that is assuming that the quantum description is incomplete. Then QM might
be completed with the addition of new ``hidden'' variables, so that the
different values of these variables would be the origin of the different
results in measurements performed in (apparently) identical conditions, as
far as we may know or control. For instance \textit{a pure state} is
represented by a wavefunction $\psi $ in standard quantum mechanics, which
means that $\psi $ contains all possible information about the state of the
system. This position had early support by Niels Bohr and his followers (the
Copenhagen interpretation of QM). In contrast an example of hidden variables
theory would represent the state by the pair $\left\{ \psi ,\lambda \right\} 
$ $,$ $\lambda $ being a random (hidden) variable. Then different results in
a measurement on a state described in QM by $\psi $ would correspond to
different values of $\lambda .$ The probabilty distribution of $\lambda $
should induce a probability distribution of the resulst of a measurement in
the state. The possibility of hidden variables was proposed, by Max Born and
other people, soon after the birth of quantum mechanics but it was rejected
by most the physicists community. After a long controversy, see e.g. \cite
{Handbook}, John Bell proved in 1965 that hidden variables are always
possible, but they cannot be local in general \cite{Bell}. Actually there
are authors critical with standard QM which do not like the solution to the
lack of causality provided by hidden variables, but demand a more radical
change of QM. It seems that this was the position of Einstein.

In summary the name ``hidden variables'' has a meaning in reference to
quantum mechanics, or more properly, to the standard (Hilbert space)
formalism of QM. In contrast ``local realism'' is a general view about
physics which considers essential that any theory should provide a
description of physical reality (realism) and the theory should preserve
relativistic causality (locality) too.

Bell \cite{Bell64} defined ``local hidden variables models'' of an
experiment those models where the results of all correlation measurements
amongst two parties, Alice and Bob, may be interpreted according to the
formulas

\begin{eqnarray}
\left\langle \theta \right\rangle &=&\int \rho \left( \lambda \right)
d\lambda M_{A}\left( \lambda ,\theta \right) ,\left\langle \phi
\right\rangle =\int \rho \left( \lambda \right) d\lambda M_{B}\left( \lambda
,\phi \right) ,  \nonumber \\
\left\langle \theta \phi \right\rangle &=&\int \rho \left( \lambda \right)
d\lambda M_{A}\left( \lambda ,\theta \right) M_{B}\left( \lambda ,\phi
\right) ,  \label{bell}
\end{eqnarray}
with standard notation. The quantities $\left\langle \theta \right\rangle
,\left\langle \phi \right\rangle $ and $\left\langle \theta \phi
\right\rangle $ are the expectation values for the results of measuring the
observables $\theta ,\phi $ and their product $\theta \phi $, respectively.
In this article I will consider that the observables correspond to
detection, or not, of some signals (``photons'') by either Alice or Bob,
attaching the values 1 or 0 to those two possibilities. In this case $%
\left\langle \theta \right\rangle ,\left\langle \phi \right\rangle $ refer
to the single, and $\left\langle \theta \phi \right\rangle $ to the
coincidence, detection probabilities in a trial of the experiment, all
trials made in identical conditions. The reported rates are the products of
these probabilities times the rate of trials. The following mathematical
conditions were assumed by Bell 
\begin{equation}
\rho \left( \lambda \right) \geq 0,\int \rho \left( \lambda \right) d\lambda
=1,M_{A}\left( \lambda ,\theta \right) \in \left[ 0,1\right] ,M_{B}\left(
\lambda ,\phi \right) \in \left[ 0,1\right] .\smallskip  \label{bell1}
\end{equation}

A constraint of locality is also included, namely $M_{A}\left( \lambda
,\theta \right) $ should be independent of the choice of the observable $%
\phi $, $M_{B}\left( \lambda ,\phi \right) $ independent of $\theta $ and $%
\rho \left( \lambda \right) $ independent of both $\theta $ and $\phi $.
From these conditions it is possible to derive empirically testable Bell
inequalities, e.g. the Clauser and Horne \cite{CH} inequality exhibited
below. The tests are most relevant if the measurements, performed by Alice
and Bob respectively, are spacially separated in the sense of relativity
theory. Indeed in this case the violation of a Bell inequality is believed
to prove that there are nonlocal influences between Alice and Bob
measurements, which would violate relativistic causality (i. e. the
assumption that no information may travel with superluminal velocity).

For experiments measuring polarization correlation of photon pairs the
Clauser-Horne (CH) inequality \cite{CH} may be written

\begin{equation}
\left\langle \theta _{1}\right\rangle +\left\langle \phi _{1}\right\rangle
+\left\langle \theta _{2}\phi _{2}\right\rangle -\left\langle \theta
_{1}\phi _{1}\right\rangle -\left\langle \theta _{1}\phi _{2}\right\rangle
-\left\langle \theta _{2}\phi _{1}\right\rangle \geq 0,  \label{CH}
\end{equation}
where $\theta _{j}$ stands for the observable ``detection by the Alice
detector put after a polarization analyzer at angle $\theta _{j}$''.
Similarly $\phi _{k}$ for Bob detector. The proof of eq.$\left( \ref{CH}%
\right) $ is well known but it is so simple that I will exhibit it in the
following. In fact if we have four real numbers fulfilling 
\begin{equation}
\theta _{1},\phi _{1},\theta _{2},\phi _{2}\in \left\{ 0,1\right\} ,
\label{angle}
\end{equation}
a trivial mathematical result is that the following inequality holds true 
\[
\theta _{1}+\phi _{1}+\theta _{2}\phi _{2}\geq \theta _{1}\phi _{1}+\theta
_{1}\phi _{2}+\theta _{2}\phi _{1}. 
\]
Taking eqs.$\left( \ref{bell1}\right) $ into account a similar inequality is
fulfilled by the response functions, that is 
\begin{eqnarray*}
&&M_{A}\left( \lambda ,\theta _{1}\right) +M_{A}\left( \lambda ,\phi
_{1}\right) +M_{A}\left( \lambda ,\theta _{2}\right) M_{A}\left( \lambda
,\phi _{2}\right) \\
&\geq &M_{A}\left( \lambda ,\theta _{1}\right) M_{A}\left( \lambda ,\phi
_{1}\right) +M_{A}\left( \lambda ,\theta _{1}\right) M_{A}\left( \lambda
,\phi _{2}\right) +M_{A}\left( \lambda ,\theta _{2}\right) M_{A}\left(
\lambda ,\phi _{1}\right) .
\end{eqnarray*}
Multiplication times the positive definite quantity $\rho \left( \lambda
\right) $ followed by integration with respect to $\lambda $ gives eq.$%
\left( \ref{CH}\right) .$ For generality and later convenience I shall write
the CH inequality in terms of the single $P_{s}$ and the coincidence $P_{c}$
probabilities as follows 
\begin{equation}
CH=P_{s}-P_{c}\geq 0,\quad 
P_{s}\equiv P_{A}+P_{B},\quad 
P_{c}\equiv P_{AB}+P_{AB^{\prime }}+P_{A^{\prime }B}-P_{A^{\prime }B^{\prime }},
\label{05}
\end{equation}
where $A$ and $A^{\prime }$ are observables for Alice and similarly $B$ and\ 
$B^{\prime }$ for Bob, all having possible values $\left\{ 0,1\right\} .$

Another popular inequality is CHSH \cite{CHSH}, which is mathematicaly
equivalent to eq.$\left( \ref{CH}\right) .$ In fact the change of variables 
\[
\theta _{1}=\frac{a_{1}+1}{2},\theta _{2}=\frac{a_{2}+1}{2},\phi _{1}=\frac{%
b_{1}+1}{2},\phi _{1}=\frac{b_{1}+1}{2} 
\]
inserted in eq.$\left( \ref{CH}\right) $ leads to the CHSH inequality 
\begin{equation}
\left\langle a_{1}b_{1}\right\rangle +\left\langle a_{1}b_{2}\right\rangle
+\left\langle a_{2}b_{1}\right\rangle -\left\langle a_{2}b_{2}\right\rangle
\leq 2,  \label{CHSH}
\end{equation}
where $a_{j},b_{j}\in \left\{ -1,1\right\} .$ In photon experiments as
discussed above CHSH might ascribe the value $1$ to detection and $-1$ to
non-detecton, rather than $\left\{ 1,0\right\} $ as CH makes. However the
equivalence arises only when both Alice and Bob measure dichotomic
observables, typically detection or non-detection of photons (see e.g. eq.$%
\left( \ref{angle}\right) ).$ In the test of CH this is almost compulsory,
but the CHSH has been used for observables with 3 possible results, rather
than 2. For instance in experiments where Alice and Bob have 2 detectors D1
and D2 each. Then the value 1 was ascribed to detection in D1, the value -1
to detection in D2, and non-detection was ignored. In these experiments CHSH
is not a true Bell inequality able to test local hidden-variables (or local
realism) except if the detectors had 100\% efficiency. In that case all
photons arriving at Alice (or Bob) would be detected and the observables
become dichotomic, whence CHSH could be used as well as CH. However in the
entangled-photons experiments performed before high-efficiency photon
detectors were available tests of CHSH required additional hypotheses. A
common assumption was that the sample of detected photons is representative
of the ensemble of all photons arriving at Alice and Bob, which is named
``fair sampling assumption''. But this extra hypothesis gave rise to a
persistent loophole in Bell tests until experiments with high-efficiency
detectors were possible \cite{Shalm}, \cite{Giustina}.

The implication eqs.$\left( \ref{bell}\right) $ $\Rightarrow $ eq.$\left( 
\ref{CH}\right) $ is a straightforward mathematical result that I will label
``Bell's theorem'' in this paper. Sometimes the name Bell's theorem is
used for the implication ``local realism'' $\Rightarrow $ eq.$\left( \ref{CH}%
\right) $ which is not a mere mathematical statement to be labeled theorem.
Furthermore the alleged implication is questionable as I shall comment on
below.

In this article I am concerned with the relation of Bell's work with
local realism, but in recent times other Bell-type inequalities have been
introduced unrelated to locality, which have relevance in quantum
information theory. Indeed technologies like quantum cryptography and
quantum computing rest on the assumption that they derive from specifically
quantum properties that could not be predicted from classical laws. Then
Bell inequalities provide a method in order to discriminate between
classical and quantum phenomena, the latter usually involving entanglement,
see e.g. \cite{Vedral}. Also similar inequalities have been used for
classical problems, for instance the study of optical coherence \cite{Saleh}%
, specially in lasers \cite{Borges}, \cite{Pereira}.

\subsection{Criticism to Bell's definition of local realism}

For many physicists the impossibility of realistic local models violating eq.%
$\left( \ref{05}\right) $ look uncontroversial. Nevertheless there is a weak
point in the conditions eqs.$\left( \ref{bell}\right) $ and $\left( \ref
{bell1}\right) $ if stated for local realism, not just for local hidden
variables. The problem does not rest on the function $\rho \left( \lambda
\right) ,$ which is quite general, but on the response functions $%
M_{A}(\lambda ,\theta )$ and $M_{B}\left( \lambda ,\phi \right) .$ For
instance $M_{A}(\lambda ,\theta )$ means that ``if we measure a dichotomic
observable (e.g. detection or non-detection) on a signal arriving at Alice
the response is \textit{yes} for some values of $\lambda $ and \textit{no}
for other values''. This statement is a typical example of quantum
measurement theory, therefore it might be appropriate for hidden variables
theories, which do not pretend to modify the quantum formalism but to
complete it with new variables. However testing just whether standard QM may
be supplemented with local hidden variables is not too relevant. In fact
neither Bohr and his followers nor Einstein liked hidden variables, although
for different reasons. Therefore in this paper I will examine whether
experiments may refute classical models. I think that the original Bell
response functions $M_{A}(\lambda ,\theta )$ and $M_{B}(\lambda ,\phi )$ are
not appropriate for the desired test.

In particular for optical experiments the functions $M(\lambda ,\theta )$
refer to photocounting, which in standard quantum mechanics is interpreted
as detection of a photon, treated as an instantaneous and simple event.
However photodetection is a complex process involving the interaction of the
incoming radiation with some resonant material system inside the detector.
This is well known in macroscopic detection, like radio or TV where the
frequency tuning is crucial. But even in quantum theory a correct analysis
of photon detection should consist of studying the evolution equations (e.g.
Schr\"{o}dinger's) of the coupled quantized electromagnetic field with
matter, during some finite time (not zero). This is for instance the case
when photon detection involves the photoelectric effect.

In order to get an appropriate substitute for the functions $M_{A}(\lambda
,\theta )$ and $M_{B}\left( \lambda ,\phi \right) $ let us study the case of
pulsed optical tests of Bell inequalities, when every pulse consists of one
time-window of duration $T$. The \textit{detection process should not be
taken as instantaneous} but as requiring a time interval $T_{\det }$ large
in comparison with the period of the incoming radiation, that is $T_{\det
}>>\omega ^{-1}.$ E.g. if $T_{\det }$ is of order nanoseconds, it is about $%
10^7$ times the period of the typical light involved. In addition
photocounting requires frequency selection in order to minimize noise.
Consequently the responses of the photodetectors should depend on the field,
including its phase, whence the alternative to Bell functions like $%
M_{A}(\lambda ,\theta )$ should be written as follows 
\begin{equation}
M_{A}\left( E_{A}\left[ \lambda ,t\right] ,\theta \right) \text{ and }%
M_{B}(\left[ E_{B}\left( \lambda ,t\right) \right] ,\phi ),  \label{07}
\end{equation}
where $E_{A}\left[ \lambda ,t\right] $ and $E_{B}\left[ \lambda ,t\right] $
are the ``amplitudes of the light arriving at Alice and Bob detectors
respectively, during a time interval $\left( t_{0},t_{0}+T\right) ".$ (In
mathematical language eqs.$\left( \ref{07}\right) $ are named \textit{%
functionals} rather than functions. This means that $M$ depends on the whole
function $E\left( t\right) ,$ not just on the value at a time). Obviously
the responses $M_{A}$ and $M_{B}$, eqs.$\left( \ref{07}\right) ,$ will
depend also on the observables, $\theta $ and $\phi $, to be measured.

Now the relevant question is whether deriving the Bell inequalities is
possible from eqs.$\left( \ref{bell}\right) $, but with eqs.$\left( \ref{07}%
\right) $ substituted for $M_{A}$ and $M_{B}$ of eqs.$\left( \ref{bell1}%
\right) .$ Proving that the answer is negative would be difficult, but
actual derivation of correlation constraints replacing Bell inequalities
would be also hard. Then I will try to answer an almost equivalent question,
that is whether there are local realistic models able to violate Bell
inequalities, like eq.$\left( \ref{CH}\right) ,$ but taking the finite
response time of detectors into account as in eqs.$\left( \ref{07}\right) .$

\subsection{Multiple detection in a trial}

There are two main differences between the functions $M_{A}$ and $M_{B}$ of
eqs.$\left( \ref{bell}\right) $ and eqs.$\left( \ref{07}\right) .$ That is
the dependence on a \textit{finite time} and the substitution of field 
\textit{amplitudes, E, for} \textit{intensities, }$I=\left| E\right| ^{2}$.
In this paper I will be concerned with intensities ignoring the phases,
which might be relevant in a more complete analysis.

As commented above in a local realistic analysis (e.g. classical)
photocounting involves the resonant transfer of the energy of a light beam
to an appropriate material device. For instance a classical detector may be
manufactured so that a detection event (a shot) is produced when the energy
transferred from the field to the detector surpasses some threshold. After a
detection event there would be a dead time such that further shots are not
possible. It is plausible to assume that the probability of a photocount in
a time interval $T$ is small for a short interval, whence $T$ cannot be too
brief in reliable experiments. In the following I will attempt getting a
mathematical model for the response function eqs.$\left( \ref{07}\right) $
which is both faithful and simple. In order to have a clue I will start
describing the plausible behaviour of the CH (Bell) inequality, eq.$\left( 
\ref{05}\right) ,$ as a function of the duration of the time-window, which
is made in the following.

If the beam intensity is weak and/or the time-window is short the
probability of a shot (the single detection probability) will be small, say $%
p<<1$ and the coincidence probability would be smaller, of order $p^{2}<<p.$
Hence in eq.$\left( \ref{05}\right) $ we will have $P_{s}\sim 2p$ and $%
P_{c}\sim $ $2p^{2}<<2p$ so that the CH (Bell) inequality will be fulfilled.
Now for given beam intensities the increase of the time-window would produce
an increase of both the single and the coincidence detection probabilities.
For instance, if the single detection probability increases from $p$ to $%
np,n>1$, then $P_{s}\sim 2np$ and $P_{c}\sim 2n^{2}p^{2},$ whence $CH\sim
2np\left( 1-np\right) ,$ which would become negative for large enough $n$,
thus violating the CH inequality. However there is a constraint, namely the
single detection probability cannot surpass unity, therefore the previous
argument is not conclusive. In the limit of very long time-window it is
plausible that photocounts are produced for any beam intensity. Then $%
p\rightarrow 1$ in eq.$\left( \ref{05}\right) $ whence we would get $%
P_{s}=P_{c}=2\Rightarrow $ $CH=0$ and the CH inequality again holds true.
The question is whether a wise choice of the angles eq.$\left( \ref{angles}%
\right) $ might produce a violation of the CH inequality, that is $CH<0$ for
some choice of the time-window neither too short nor too long. Now our goal
will be to construct a local model able to produce the violation of the CH
inequality eq.$\left( \ref{05}\right) $.

In order to get a model I shall start finding the response functions eqs.$%
\left( \ref{07}\right) .$ If we assume that within a time-window either one
shot or none can be produced (but not many), then the time $t$ will not
enter in eqs.$\left( \ref{07}\right) $ and we will arrive at Bell's
proposal, eq.$\left( \ref{bell}\right) ,$ whence any local realistic model
will fulfil the Bell inequalities. An escape to that situation is to
consider the possibility of more than one shot within the time window and
the most simple model for eqs.$\left( \ref{07}\right) $ is to assume that
there may be either 0, 1 or 2 shots. In this case there may be up to $%
2\times 2=4$ coincidence detections, whence we might have $P_{c}>P_{s}$ in
eq.$\left( \ref{05}\right) $ thus violating the CH (Bell) inequality. In the
following I shall obtain the probabilities of the possible results in a
particular but interesting case.

Let $T$ be the duration of the time-window chosen for a trial of the
experiment, and let $p_{A}\in \left[ 0,1\right] $ be the probability of at
least one detection event by Alice in the first half time-window with
duration T/2 and also $p_{A}$ for the second half. Then the probability of
no counts by Alice in the whole window will be $(1-p_{A})^{2}$ assuming that
both events (of no detection) are statistically independent. Hence the
probability of at least one count by Alice during the time window T will be 
\begin{equation}
P_{A}=1-\left( 1-p_{A}\right) ^{2}=2p_{A}-p_{A}^{2}=p_{A}\left(
2-p_{A}\right) =P_{B}.  \label{p1}
\end{equation}
Similar for Bob assuming $p_{B}=p_{A}$.

Finding the probability of coincidence detection in the whole window is more
involved. There are four possible individual detection events, namely A1,
A2, B1, B2, where A and B stand for Alice and Bob detectors, respectively,
and the numbers 1, 2 stand for at least one detection in the first and the
second half window respectively. I shall report a coincidence whenever one
or several amongst the following composite events takes place: 
\begin{equation}
A1\frown B1,A1\frown B2,A2\frown B1,A2\frown B2,  \label{p3}
\end{equation}
where $\frown $ means logical union. For instance $A1\frown B2$ means that
there is a count by Alice in the first half time window and another count by
Bob in the second half. This with or without additional counts in either $A2$
or $B1,$ or in both. The probability of this composite event is 
\[
\Pr \left( A1\frown B2\right) =\Pr \left( A1\right) \times \Pr (B2/A1) 
\]
where $B2/A1$ means event $B2$ conditional to event $A1$. I assume that
events in different half windows are independent whence 
\[
\Pr (B1/A2)=\Pr (A2)=p_{A}=\Pr (B2/A1). 
\]
However events in the same half window are positively correlated, that is 
\[
\Pr (B1/A1)=\Pr (B2/A2)=q_{B}>p_{A},q_{B}\leq 1. 
\]
Indeed the experiments are usually prepared so that $q_{B}$ is large. Then
we have 
\begin{eqnarray}
\Pr \left( A1\frown B2\right) &=&\Pr \left( A2\frown B1\right) =p_{A}p_{B}, 
\nonumber \\
\Pr \left( A1\frown B1\right) &=&\Pr \left( A2\frown B2\right)
=p_{A}q_{B}=p_{B}q_{A}.  \label{30}
\end{eqnarray}
I stress that in eq.$\left( \ref{30}\right) $ $p_{A}p_{B}$ correspond to
uncorrelated events but $p_{A}q_{B}$ or $p_{B}q_{A}$ to correlated events.
For instance a classical chaotic ligth the intensity $I(t)$ is fluctuating,
that is for some times the intensity will be large and for other times
small. The intensities arriving at Alice$,$ $I_{A}(t)$, and Bob$,$ $I_{B}(t)$%
, will be correlated if both the high and the low values of the intensity
appear simultaneously in $I_{A}$ and $I_{B},$ the maximal correlation
corresponding to $I_{A}(t)=I_{B}(t).$ Thus we might name $p_{B}q_{A}$ the
probability of a fair coincidence and $p_{A}p_{B}$ probability of an
accidental coincidence. The distinction will be relevant in section 3.3.

The negations of the events eq.$\left( \ref{p3}\right) $ are exclusive and
independent, whence the probability that no coincidence detections takes
place in the whole window is 
\[
\Pr \left( \text{no coincidences}\right) =\prod_{j,l=1,2}\left[ 1-\Pr \left(
Aj\frown Bl\right) \right] =\left( 1-p_{A}p_{B}\right) ^{2}\left(
1-p_{A}q_{B}\right) ^{2}, 
\]
whence 
\begin{equation}
\Pr \left( \text{at least one coincidence}\right) \equiv P_{AB}=1-\left(
1-p_{A}p_{B}\right) ^{2}\left( 1-p_{A}q_{B}\right) ^{2}.  \label{p2}
\end{equation}

Let us compare the ratio $P_{AB}/P_{A}=P_{AB}/P_{B}$ with the ratio $\left(
pq\right) /p=q,$ for the simple case when $p_{A}=p_{B}$ and $%
q_{A}p_{B}=p_{A}q_{B},$ corresponding to the first half window alone. Taking
eqs.$\left( \ref{p1}\right) $ and $\left( \ref{p2}\right) $ into account, we
get 
\begin{equation}
gain=\frac{1-(1-p^{2})^{2}(1-pq)^{2}}{1-(1-p)^{2}}-q.  \label{gain}
\end{equation}

The quantity \textit{gain} is positive for all possible values of $p\in
\left[ 0,1\right] $ and $q\in \left[ p,1\right] $. \textit{Proof}: For $q=0$
we easily get 
\[
gain=\frac{1-(1-p^{2})^{2}}{1-(1-p)^{2}}>0. 
\]
For any $q\neq 0$ the derivative fulfils 
\begin{eqnarray*}
\frac{d\left( gain\right) }{dq} &=&\frac{1+(1-p^{2})^{2}2(1-pq)p}{1-(1-p)^{2}%
}-1 \\
&=&\frac{(1-p^{2})^{2}2(1-pq)p+(1-p)^{2}}{1-(1-p)^{2}}>0,
\end{eqnarray*}
where in the latter inequality we have taken into account that the
coincidence detection probability $pq\leq 1.$ Then $gain>0.$

We conclude that the ratio between coincidences and singles may \textit{%
increase with the size of the time window}. In the CH inequality, eq.$\left( 
\ref{05}\right) ,$ single detections enter with positive sign and
coincidence detections mainly with negative sign. Hence the result $gain>0$
suggests that the probability increases of violating the CH inequality, eq.$%
\left( \ref{05}\right) $, if we just record whether there are detections, or
not, by Alice and similar for Bob, that is without specifying how many shots
have been produced, if any.

It is the case that the same happens if the dead time of the detector is
long enough to prevent more than one shot within a window. In fact the
probability of \textit{zero} detection events in the whole time-window is $%
\left( 1-p_{A}\right) ^{2},$ see eq.$\left( \ref{p1}\right) $, either if
there is a dead time or not, because the dead time may change just the
probability of the \textit{second }shot, which the dead time may prevent.
Thus the probability of ``at least one shot'' with no dead time is the same
as the probability of just one shot when there is a long dead time.
Similarly for Bob detections. The conclusion is that the result eq.$\left( 
\ref{gain}\right) $ remains the same even if the set-up of the experiment is
able to discriminate between one shot and several within a time-window.

\subsection{The coincidence-time loophole}

The possibility that a long time window would give rise to a fake violation
of the Bell inequality by a local model was pointed out by Larsson and Gill
(the so called ``time-coincidence loophole'') \cite{Larsson}. The loophole
was claimed to be closed by Ag\"{u}ero et al. \cite{Aguero} in a pulsed
set-up in which the time coincidence window and delay could be changed at
will, even after the experiment's end. The argument \cite{Aguero} is that
in experiments with entangle photon pairs it is possible to know whether
coicidence counts correspond to the same photon pair, then considered fair
coincidences, or to accidental (spureous) coincidences due to other causes.
That is the experimental discrimination is assumed possible via recording
precisely the time of detection. The loophole is closed if coincidences for
the test of the Bell inequality are discarded whenever there is no guarantee
that the coincidence detection corresponds to two photons from the same
entangled pair.

An analysis of the coincidence-time loophole has been made by Christensen et
al. \cite{Kwiat}. In the article several experiments with classical ligth
are reported, and the analysis of them made, which shows that for small
time-windows the Bell inequalities hold true, but with long windows there
are (fake) violations. The classical experiments were made with laser light 
\cite{Kwiat}, but models with chaotic light would be worthwhile. In fact the
classical field fluctuations mentioned after eq.$\left( \ref{30}\right) $
may simulate quantum entanglement, which is not the case for laser light.
The argument may be clarified with the following simplistic model of two
classical correlated light beams.

Firstly I will investigate the distribution of photocounts in a detector
along the time. I may assume that the detection probability, \textit{P}$%
\left( t\right) $, at a given time $t$ is proportional to the ligth
intensity $I(t)$ arriving at the detector at that time. A more physical
hypothesis would be to restrict the intensity to a narrrow range of
frequencies (those in resonance with the detector) and also to assume that
the detection probability is proportional to that intensity integrated
during some detection time interval $\left( t-T_{\det },t\right) $ rather
than the instantaneous intensity. For simplicity I shall assume $P\left(
t\right) \varpropto I\left( t\right) ,$ which may be valid if $T_{\det }$ is
much smaller than the time-window. Let us start with the case of
monochromatic light where the wave field amplitude, $\phi ,$ and the
intensity, $I$, may be written 
\begin{equation}
\phi (t)=A\exp (i\nu t+i\mathbf{k\cdot x}),\left| \mathbf{k}\right| =\nu
/c,I(t)=\left| \phi (t)\right| ^{2}=\left| A\right| ^{2}=\text{constant.}
\label{d0}
\end{equation}
Eq.$\left( \ref{d0}\right) $ represents a plane wave (extended over the
whole space) but in experiments we need beams (localized in the transverse
directions) and this requires the interference of many plane waves. In the
following I shall consider waves at a given space point which I take as the
origin of the spatial coordinate system, that is $\mathbf{x=0}$. Then I may
ignore the space dependence.

Let us consider the case of just two plane waves with frequencies $\omega $
and $\nu $ with similar amplitude $A$, so that the field may be written 
\[
\phi (t)=A\exp \left( i\omega t\right) +A\exp \left( i\nu t\right) =2A\cos
\left( \frac{\omega -\nu }{2}t\right) \exp \left( i\frac{\omega +\nu }{2}%
t\right) , 
\]
which might be considered a single wave with frequency $\frac{\omega +\nu }{2%
}$ and modulated amplitude. The point is that modulation is unavoidable in
classical light beams and it gives rise to intensity fluctuations. For
instance in our example the intensity changes with time as follows 
\[
I=4\left| A\right| ^{2}\cos ^{2}\left( \frac{\omega -\nu }{2}t\right) . 
\]

Intensity fluctuations may be small in the case of laser beams, but may be
large in more common (chaotic or thermal) light. For instance let us
consider a wave which is the superposition of 3 field amplitudes as follows 
\begin{equation}
\phi (t)=a\cos \left( \omega t\right) -2a\cos \left( 2\omega t\right) +a\cos
\left( 3\omega t\right) ,a=A\exp (i\nu t).  \label{d1}
\end{equation}
It is trivial to get explicitly the intensity as a function of $t$, but for
our argument we need only the mean intensity, which is $6\left| A\right|
^{2}/2\pi \simeq 0.96\left| A\right| ^{2},$ and the maximal intensity, which
is $16\left| A\right| ^{2}$ and it occurs at times $t=\left( 2n+1\right) \pi
/\omega ,$ $n$ being an integer.

Now let us consider two almost monochromatic light beams corresponding to
two lasers (with weak intensity fluctuations), roughly represented by eq.$%
\left( \ref{d0}\right) $ each, that arrive at Alice and Bob respectively. If
we assume that the single detection probability (within a time-window) is
proportional to the arriving intensity, then the coincidence probability
will be the product of the intensities $I_{A}\left( t\right) $ and $%
I_{B}\left( t^{\prime }\right) $ at Alice and Bob respectively, that is $%
I_{A}I_{B}$ independent of the time interval $t^{\prime }-t.$ This implies
that the detection times by Alice and Bob are uncorrelated, in sharp
contrast with the simultaneous detection times ascribed to entangled photon
pairs. In fact let us assume that the rate of detection events by either
Alice or Bob is $1/T$, then the typical time elapsed between a detection by
Alice (say a shot) and the most close shot by Bob will be of order $T/4$.
This explains why the coincidences with short time delay $t^{\prime }-t$ are
scarce in the experiment by Christensen et al. $\cite{Kwiat}$ , thus
fulfilling the Bell inequality if only coincidences with short delay are
taken as genuine.

In sharp contrast if the beams arriving at Alice and Bob were both of the
form eq.$\left( \ref{d1}\right) ,$then the product of intensities $I_{A}I_{B}
$ at times $t\simeq t^{\prime }\simeq \left( 2n+1\right) \pi /\omega $ is
about $\left( 16/0.96\right) ^{2}\simeq 280$ times greater than the average.
The result is that the coincidences will be much more probable at times
around $\left( 2n+1\right) \pi /\omega $, and also the time interval $\left|
t-t^{\prime }\right| $ elapsed between Alice and Bob detections would be
likely short.

Also it is worth mentioning that spontaneous parametric down conversion may
be interpreted as the interaction of both, the pumping laser and the
electromagnetic quantum vacuum (zeropoint) field, with electrons in the
nonlinear crystal \cite{Kaled}, see also \cite{Universe}, \cite{FOOP25}.
These electrons are excited and then radiate in two different directions
with frequencies whose sum equals the pumping laser frequency. The process
is not instantaneous whence the coincidence detection of entangled photons
may not be strictly simultaneous.

Our classical detection model illustrates that the discrimination between
quantum coincidences (involving entangled photons) and classical
coincidences (involving correlated fluctuating intensities) may not be easy.
Indeed a time interval T is unknown, greater than any time lapse for
entangled photons coincidence detection but smaller than any (allegedly)
spureous coincidence with classical light. Thus claiming that the
coincidence-time loophole has been closed is questionable in my opinion.

\section{Analysis of a realistic local model}

In this section I will propose a model resting on classical electrodynamics
which is able to violate a Bell inequality. This will provide a
counter-example in order to prove that the empirical violation of a Bell
inequality does not refute local realism. I stress that I shall not try to
analyze actually performed experiments, but to contrive a model for a
typical experiment. I will present the model in section 3.1, and in 3.2 I
will show that a standard analsis of the model give predictions fulfilling
the CH inequality eq.$\left( \ref{05}\right) .$ In section 3.3 I shall
analyze the model with the modifications proposed in section 2.4.

\subsection{The model}

The model consists of a source that may send light beams with the same
frequency, $\omega ,$ to two parties (Alice and Bob). Treating the beams in
terms of the amplitudes, those sent to Alice and Bob in a trial have vector
amplitudes 
\begin{equation}
\vec{E}_{Alice}=a\vec{H}+b\vec{V},\vec{E}_{Bob}=b\vec{H}+a\vec{V},
\label{00}
\end{equation}
respectively, $\vec{H}$ and $\vec{V}$ being unit vectors in the horizontal
and vertical directions, respectively. I assume that both fields have the
same frequency, which may be represented as $a\left( t\right) \cos (wt+\chi
) $, and the amplitudes of the fields, $a\left( t\right) $ and $b\left(
t\right) $, have a slow time dependence which I will not make explicit. I
assume the following joint probability distribution $\rho \left( a,b\right) $
for the averages of the amplitudes within a time window 
\begin{equation}
\rho \varpropto \exp \left( -\left| a\right| ^{2}-\left| b\right|
^{2}\right) ,  \label{0}
\end{equation}
that is the amplitudes are Gaussian and statistically independent of each
other. The phases are random uniformly distributed in $\left( 0,2\pi \right) 
$. Thus each beam, with field amplitude $a\left( t\right) $ or $b\left(
t\right) $, has properties of chaotic light which is typical in (almost)
monochromatic beams of macroscopic physics. An appropriate set up in the
source, involving beam-splitters, mirrors and polarizers, may allow
producing, from the usual chaotic light beams with amplitudes $a$ and $b,$
another polarized beams with amplitudes $\vec{E}_{Alice}$ and $\vec{E}_{Bob}$
eq.$\left( \ref{00}\right) .$ Thus the described model may be considered
classical, and therefore realistic local.

The beam $\vec{E}_{Alice}$ arriving at Alice will cross a polarization
analizer at an angle $\theta $ with the horizontal, whence the beam reaching
the detector, put after the analizer, will have an amplitude 
\[
E_{A}=a\cos \theta +b\sin \theta , 
\]
and it shall be linearly polarized at angle $\theta .$ Similarly for the
amplitude arriving at Bob's detector, after crossing an analizer put at
angle $\phi ,$ that is 
\[
E_{B}=b\sin \phi +a\cos \phi . 
\]
The intensities are the square moduli of the amplitudes, that is 
\begin{eqnarray}
I_{A} &=&\left| a\right| ^{2}\cos ^{2}\theta +\left| b\right| ^{2}\sin
^{2}\theta +2\left| a\right| \left| b\right| \cos \chi \cos \theta \sin
\theta ,  \label{2} \\
I_{B} &=&\left| a\right| ^{2}\cos ^{2}\phi +\left| b\right| ^{2}\sin
^{2}\phi +2\left| a\right| \left| b\right| \cos \xi \cos \phi \sin \phi , 
\nonumber
\end{eqnarray}
where $\left| a\right| ^{2}$ and $\left| b\right| ^{2}$ are averaged
intensities during a time-window, $\chi $ and $\xi $ are the relative phases
of the fields $a$ and $b$ in $E_{A}$ and $E_{B}$ respectively. I shall
assume that $\chi $ and $\xi $ are statistically independent. Indeed the
phases are very sensitive to the path length from the source to the
detectors and those paths are different for Alice and Bob. In order to
complete the model I must introduce the response functions eqs.$\left( \ref
{07}\right) ,$ which will be postponed to sections 3.2 and 3.3.

\subsection{Standard analysis}

For the sake of clarity and later convenience I shall devote this subsection
to a standard analysis of the model, that is: 1) I will ignore the phases,
2) I assume that the detection probability is a function of the average
light intensity arriving at the detector within a time-window, and 3) I
assume that within a time-window there may be one detection event (one shot)
or none, that is we cannot have 2 or more shots. For simplicity the
detection function $p\left( I\right) $ is chosen the same for both Alice and
Bob. That is $p(I_{A})$ will be the probability that there is one shot in
Alice's detector during a time-window. Similarly $p(I_{B})$ for Bob. The
probability $p_{c}$ of one coincidence detection within the window will be 
\[
p_{c}=\int \rho \left( \left| a\right| ,\left| b\right| \right)
p(I_{A})\times p(I_{B})d\left| a\right| ^{2}d\left| b\right| ^{2}, 
\]
where $\rho \left( \left| a\right| ,\left| b\right| \right) $ is the
probability distribution eq.$\left( \ref{0}\right) .$

The funcion $p\left( I\right) $ plays the role of $M_{A}\left( \lambda
,\theta \right) $ and $M_{B}\left( \lambda ,\phi \right) $ in eqs.$\left( 
\ref{bell}\right) ,$ the angles $\left\{ \theta ,\phi \right\} $ being
included in the intensities eqs.$\left( \ref{2}\right) .$ The function $%
p\left( I\right) ,$ being a probability, should fulfil the constraint 
\begin{equation}
0\leq p\left( I\right) \leq 1.  \label{1}
\end{equation}
In order to choose $p\left( I\right) $ I consider that it is likely small or
zero for small intensity and it remains below unity for any intensity. I
propose 
\begin{equation}
p\left( I\right) =1-\exp \left( -kI\right) ,  \label{4}
\end{equation}
where $k>0$ is a constant, the same for both Alice's and Bob's
detectors. The probability function eq.$\left( \ref{2}\right) $ is fairly
proportional to the intensity for small $I,$ then increases but fulfilling
the constraint eq.$\left( \ref{1}\right) $ for any intensity.

The following property of the detection probability eq.$\left( \ref{4}%
\right) $ is interesting. If we assumed that the detection probability
within a short time interval with duration $T/n$ is $\varepsilon I$, $T$
being the time-window, then the probability of non-detection in this short
time interval will be $1-\varepsilon I,$ and detection in none of the short
time intervals $\left( 1-\varepsilon I\right) ^{n}$ $\simeq \exp \left(
-n\varepsilon I\right) .$ Puting $k\equiv $ $n\varepsilon $ leads to eq.$%
\left( \ref{4}\right) .$ However I do not choose the function eq.$\left( \ref
{4}\right) $ for that property in our contrived model, but for calculation
convenience.

In eq.$\left( \ref{0}\right) $ the exponent should be dimensionless, which
implies that I ignore a dimensional parameter. This also means that the
parameter $k$ of eq.$\left( \ref{4}\right) $, which is the inverse of a
field intensity, is also treated as dimensionless, and this will be also the
case in the rest of the article. Putting appropriate dimensional parameters
would be straightforward but give rise to more involved notation.

The single detection probabilities by Alice and Bob are given by eqs.$\left( 
\ref{2}\right) $ without the phase factor (which is equivalent to putting $%
\chi =\xi =\pi /2).$ In order to simplify the notation I will introduce new
variables

\begin{equation}
x\equiv \left| a\right| ^{2},y\equiv \left| b\right| ^{2}.  \label{10}
\end{equation}
Then the single detection probability by Alice will be, taking eqs.$\left( 
\ref{0}\right) $ to $\left( \ref{4}\right) $ into account, 
\begin{eqnarray}
P_{A} &=&\int_{0}^{\infty }d\left| a\right| ^{2}\int_{0}^{\infty }d\left|
b\right| ^{2}\exp \left( -\left| a\right| ^{2}-\left| b\right| ^{2}\right)
\left[ 1-\exp \left( -kI_{A}\right) \right]  \nonumber \\
&=&1-\int_{0}^{\infty }dx\int_{0}^{\infty }dy\exp \left( -x-y\right) \exp
\left( -kx\cos ^{2}\theta -ky\sin ^{2}\theta \right) .  \label{5}
\end{eqnarray}
The integrals are trivial and we get 
\begin{equation}
P_{A}=1-Q_{A},Q_{A}=\frac{1}{(1+kc^{2})(1+ks^{2})}=\frac{1}{%
1+k+k^{2}c^{2}s^{2}},  \label{6}
\end{equation}
where $c\equiv \cos \theta ,s\equiv \sin \theta .$ Similarly the single
detection probability by Bob is 
\begin{equation}
P_{B}=1-Q_{B},Q_{B}=\frac{1}{1+k+k^{2}c^{\prime 2}s^{\prime 2}},  \label{6a}
\end{equation}
with $c^{\prime }\equiv \cos \phi ,s^{\prime }\equiv \sin \phi .$

The coincidence probability will be 
\begin{eqnarray}
P_{AB} &=&\int_{0}^{\infty }dx\int_{0}^{\infty }dy\exp \left( -x-y\right)
\left[ 1-\exp \left( -kI_{A}\right) \right] \left[ 1-\exp \left(
-kI_{B}\right) \right]  \nonumber \\
&=&P_{A}+P_{B}-1+Q_{AB},  \label{7}
\end{eqnarray}
where 
\begin{eqnarray}
Q_{AB} &\equiv &\int_{0}^{\infty }dx\int_{0}^{\infty }dy\exp \left(
-x-y\right) \exp \left( -kI_{A}-kI_{B}\right)  \nonumber \\
&=&\int_{0}^{\infty }dx\int_{0}^{\infty }dy\exp \left( -x-y\right) \exp
\left[ -kx\left( c^{2}+c^{\prime 2}\right) -ky\left( s^{2}+s^{\prime
2}\right) \right]  \nonumber \\
&=&\frac{1}{\left[ 1+k\left( c^{2}+c^{\prime 2}\right) \right] \left[
1+k\left( s^{2}+s^{\prime 2}\right) \right] }  \nonumber \\
&=&\frac{1}{1+2k+k^{2}\left( c^{2}+c^{\prime 2}\right) \left(
s^{2}+s^{\prime 2}\right) }.  \label{8}
\end{eqnarray}
In order to test the Clauser-Horne (Bell) inequality $\left( \ref{05}\right) 
$ I shall choose the angles 
\begin{equation}
A=\pi /6,B=\pi /3,A^{\prime }=0,B^{\prime }=\pi /2,  \label{angles}
\end{equation}

For the sake of clarity I get explicitly the values of single and
coincidence detection probability for the particular polarizer angles $A=\pi
/6,B=\pi /3$. We have 
\begin{eqnarray}
P_{A} &=&P_{B}=1-Q_{A},Q_{A}=\frac{1}{1+k+\frac{3}{16}k^{2}},  \nonumber \\
P_{AB} &=&P_{A}+P_{B}-1+Q_{AB},Q_{AB}=\frac{1}{1+k+k^{2}},  \label{11}
\end{eqnarray}
where eqs.$\left( \ref{6}\right) $ to $\left( \ref{8}\right) $ have been
taken into account.

It is convenient to write the quantity CH, eq.$\left( \ref{05}\right) ,$ as
a difference between two terms, the first one consisting of single
probabilities and the second one consisting of coincidence probabilities as
in eq.$\left( \ref{05}\right) $, that is

\[
CH=Ps-Pc,Ps=P_{A}+P_{B},Pc=P_{AB}+P_{AB^{\prime }}+P_{A^{\prime
}B}-P_{A^{\prime }B^{\prime }}. 
\]
From eqs.$\left( \ref{11}\right) $ we get 
\[
Ps=2-\frac{2}{1+k+\frac{3}{16}k^{2}}. 
\]
Similarly we may calculate the coincidence probabilities involved in $Pc$
and we get after some algebra 
\[
Pc=2-\frac{4}{1+k+\frac{3}{16}k^{2}}-\frac{2}{1+k+\frac{7}{16}k^{2}}, 
\]
which leads to 
\begin{equation}
CH=Ps-Pc\equiv \frac{2}{1+k+\frac{3}{16}k^{2}}-\frac{2}{1+k+\frac{7}{16}k^{2}%
}>0,  \label{9}
\end{equation}
The model predictions fit in the Clauser-Horne inequality for all values of $%
k>0$, as it should in view of ``Bell's theorem'', see section 2. In fact
the theorem states that for any choice of the function $p\left( I\right) $
fulfilling eq.$\left( \ref{1}\right) $ the CH inequality holds true for any
classical model (with standard analysis).

\subsection{The violation of a Bell inequality}

In this section I present a local realistic model similar to that of the
previous section, but now assuming a time-window with the possibility of two
photocounts, one during the ``first half time-window'' and another one
during the ``second half'' by Alice, and similar for Bob. This for every
pair of polarizer's angles.

I start with the case of the polarization angles $A$ for Alice and $B$ for
Bob, see eq.$\left( \ref{angles}\right) $. The probability that Alice
detector does not record any detection events in the whole time-window, that
is neither in the first half nor in the second one, will be 
\[
(1-P_{A})^{2}=Q_{A}^{2}, 
\]
where eq.$\left( \ref{6}\right) $ has been taken into account. Then the
probability of having at least one detection event (a shot) in the whole
time-window by Alice will be 
\begin{equation}
P_{A}^{m}=1-(1-P_{A})^{2}=1-Q_{A}^{2},  \label{a1}
\end{equation}
and similar for Bob 
\begin{equation}
P_{B}^{m}=1-(1-P_{B})^{2}=1-Q_{B}^{2},  \label{b1}
\end{equation}
where the superindex $m$ stands for ``multiple'' or ``modified'' detection.

The probability of no coincidence in the first half and no coincidence in
the second half will be 
\begin{equation}
(1-P_{AB})^{2},  \label{a2}
\end{equation}
the probability of one shot by Alice in the first half and one shot by Bob
in the second half is $P_{A}P_{B},$ and the probability that that this does
not happen will be 
\begin{equation}
1-P_{A}P_{B}.  \label{a3}
\end{equation}
Similarly the probability of one shot by Bob in the first half and one shot
by Alice in the second half is $P_{A}P_{B}$ and the probability that this
does not happen will again be 
\begin{equation}
1-P_{A}P_{B}.  \label{a4}
\end{equation}
Then the probability that no coincidence takes place in the whole
time-window will be the product of eqs.$\left( \ref{a2}\right) ,\left( \ref
{a3}\right) $ and $\left( \ref{a4}\right) ,$ that is $(1-P_{AB})^{2}\left(
1-P_{A}P_{B}\right) ^{2},$ whence the probability of at least one
coincidence detection will be 
\begin{eqnarray*}
P_{AB}^{m} &=&1-(1-P_{AB})^{2}\left( 1-P_{A}P_{B}\right) ^{2} \\
&=&1-\left[ 1-\left( P_{A}+P_{B}-1+Q_{AB}\right) \right] ^{2}\left[
1-(1-Q_{A})(1-Q_{B}\right] ^{2} \\
&=&1-\left[ 1-\left( 1-Q_{A}-Q_{B}+Q_{AB}\right) \right] ^{2}\left[
Q_{A}+Q_{B}-Q_{AB}\right] ^{2} \\
&=&1-\left[ Q_{A}+Q_{B}-Q_{AB}\right] ^{4}
\end{eqnarray*}
where eqs.$\left( \ref{p2}\right) $ and $\left( \ref{11}\right) $ have been
taken into account.

After some algebra it is possible to obtain the coincidence detection
probabilities for the remaining pairs of angles and then to get the Bell
inequality eq.$\left( \ref{05}\right) $ appropriate for the model with
multiple detection, that is 
\[
CH=P_{A}^{m}+P_{B}^{m}+P_{A^{\prime }B^{\prime
}}^{m}-P_{AB}^{m}-P_{AB^{\prime }}^{m}-P_{A^{\prime }B}^{m}\geq 0, 
\]
which in terms of the quantities Q reads

\begin{eqnarray*}
CH &=&\left[ 2Q_{A}-Q_{AB}\right] ^{4}+2\left[ Q_{A}+Q_{B^{\prime
}}-Q_{AB^{\prime }}\right] ^{4} \\
&&-2Q_{A}^{2}-\left[ 2Q_{A^{\prime }}-Q_{A^{\prime }B^{\prime }}\right] ^{4}
\end{eqnarray*}
where

\[
Q_{A}=\frac{1}{1+k+k^{2}c^{2}s^{2}}=\frac{1}{1+k+\frac{3}{16}k^{2}}%
,Q_{A^{\prime }}=Q_{B^{\prime }}=\frac{1}{1+k} 
\]
\[
Q_{AB}=\frac{1}{1+2k+k^{2}\left( c^{2}+c^{\prime 2}\right) \left(
s^{2}+s^{\prime 2}\right) }=\frac{1}{1+2k+k^{2}}=\frac{1}{\left( 1+k\right)
^{2}} 
\]
\[
Q_{AB^{\prime }}=Q_{A^{\prime }B}=\frac{1}{1+2k+\frac{15}{16}k^{2}}%
,Q_{A^{\prime }B^{\prime }}=\frac{1}{\left( 1+k\right) ^{2}}. 
\]

Finally we obtain 
\begin{eqnarray}
CH &=&2\left[ Q_{A}+Q_{B^{\prime }}-Q_{AB^{\prime }}\right] ^{4}-2Q_{A}^{2} 
\nonumber \\
&=&2\left[ \frac{1}{1+k}+\frac{1}{1+k+\frac{3}{16}k^{2}}-\frac{1}{1+2k+\frac{%
15}{16}k^{2}}\right] ^{4}  \nonumber \\
&&-2\left[ \frac{1}{1+k+\frac{3}{16}k^{2}}\right] ^{2}.  \label{fin}
\end{eqnarray}
This result violates the CH inequality CH $>0$ for some values of
k, as may be checked. In fact expanding this expression up to first order in 
$k$, in both the first and the second terms, leads to 
\[
CH=2\left[ 2-2k-1+2k\right] ^{4}-2\left[ 1-k\right] ^{2}+O\left(
k^{2}\right) \simeq 2-2\left( 1-2k\right) =4k>0, 
\]
a result which is a valid approximation for small $k$ and it fulfils the CH
inequality. However for large $k$ the bracket of the first term is $\left(
1+k\right) ^{-1}\simeq 1/k,$ the second one being ($16/3)k^{-2}$, whence 
\[
CH=2k^{-4}-2\left( \frac{16}{3}\right) ^{2}k^{-4}<0. 
\]
The result is that $CH>0$ for small $k$ but $CH<0$ for large $k$, violating
the CH (Bell) inequality$.$ That is $CH$ is positive for small $k$,
decreases to the value $CH=0$ for $k\simeq 1$, and it is negative for $k>1$
but increases again for large $k$ reaching $0$ in the limit $k\rightarrow
\infty .$

The analysis of a classical model is an example of the coincidence-time
loophole mentioned in section 2.3. In our analysis the (Bell) CH inequality
is violated by the model predictions if we do not distinguish whether there
is one shot or there are many by Alice in one trial of the experiment, and
similar for Bob. That is in the CH inequality we compare the rate of trials
with at least one detection by Alice, and similar for Bob, with the rate of
trials with at least one coincidence in the whole time-window, this for
every pair of polarizer angles. The violation is even more clear if we
consider a model where photocounters have a dead time so long that once a
detection event is produced in Alice detector, any further single detection
event in the same trial being inhibited for some time, and similar for Bob,
see the end of section 2.4. With this assumption the probability of ``one
shot'' is the \textit{same} than the probability of ``at least one shot''
which we have got in our analysis. Similarly for the probability of one
coincidence detection. As a conclusion our model violates the Bell
inequality if the time window is large enough so that coincidences are taken
into account even if there is some delay between the two single detections.

\section{Conclusions}

Bell's conditions for local realism and the derivation of Bell
inequalities have been critically revisited. In optical experiments I refuse
an extremely idealized relation between signals and detection probabilities,
as was proposed by Bell in eqs.$\left( \ref{bell}\right) $ and $\left( \ref
{bell1}\right) $. In contrast I have pointed out that the response to light
signals may depend in a complex manner on the interaction between signal and
detector, not just on the instantaneous intensity of the signal. In
particular I analyze the dependence of the results on the duration of the
time-windows in a pulsed experiment, which gives rise to the
coincidence-time detection loophole.

The current opinion is that the loophole has been closed via performing the
Bell experiments with small enough detection time-windows, say with duration
T. The argument is that for windows greater than T no coincidences by
photons of the same entangled pair are lost, but coincidence detection
within a time smaller than T is not possible \textit{for any} classical
beam. I believe that this procedure is disputable because the\textit{\
precise size T} of the time-window appropriate for the discrimination
between genuine and spureous coincidences is not well defined. Experiments
performed with laser light \cite{Kwiat} show that the quantity T may be
easily determined because detection events by Alice and Bob are uncorrelated
in time. However I argue that for classical (chaotic) light detection events 
\textit{may be concentrated} in periods of high intensity (positive
fluctuations). In summary I claim that the time-coincidence loophole has not
been closed, whence the empirical violation of Bell inequalities does not
refute local realism.

I point out that in order to refute classical models the analysis of the
experiments should not rely on quantum concepts as those involved in
``quantum measurement theory''. Indeed in a classical analysis there is no
distinction between fair and spureous coincidences. Also the analysis should
avoid assuming a corpuscular nature of light (photons).

A realistic local model involving light beams correlated in polarization is
exhibited where the results violate the Clauser-Horne (Bell) inequality.

\section{Statements and Declartations}

No statements and declarations required

\end{document}